\documentclass[twocolumn,amsmath,amssymb,groupedaddress,superscriptaddress]{revtex4-1}
\usepackage[bookmarks=true,colorlinks,linkcolor=blue,urlcolor=blue,citecolor=blue]{hyperref}
\usepackage{graphics,color}
\usepackage{amsmath}
\usepackage[dvips]{graphicx}
\usepackage{float,graphicx}
\usepackage{subfigure,hyperref}
\usepackage{epstopdf}
\usepackage{enumitem}
\usepackage{soul}
\usepackage{epstopdf}
\newcommand{\beq}{\begin{equation}}
\newcommand{\eeq}{\end{equation}}
\newcommand{\beqa}{\begin{eqnarray}}

\newcommand{\eeqa}{\end{eqnarray}}
\newcommand{\unit}{1\!\!1}

\newcommand{\abs}[1]{\left\vert #1\right\vert}
\def\e{\epsilon}

\def\s{\sigma}

\newcommand{\bra}[1]{\langle #1 \vert} 
\newcommand{\ket}[1]{\vert #1 \rangle} 

\def\be{\begin{equation}}
\def\ttr{\textmd{tr}}

\def\ee{\end{equation}}

\def\ttr{\textmd{tr}}

\def\Rre{\textmd{Re}}

\def\Ttr{\textmd{Tr}}

\def\be{\begin{equation}}
\def\ee{\end{equation}}
\def\w{\omega}
\def\bea{\begin{eqnarray}}
\def\eea{\end{eqnarray}}

\def\np{n^\prime}
\def\mpr{{m^\prime}}

\usepackage{bm}
\usepackage{graphicx}
\begin{document}
\title{Dephasing dynamics of an impurity coupled to an anharmonic environment}
\author{Max Bramberger}
\author{In\'es De Vega}
\affiliation{Department of Physics and Arnold Sommerfeld Center for Theoretical
Physics, Ludwig-Maximilians-University Munich, Germany}

\begin{abstract}
We analyze the dephasing dynamics of an impurity coupled to an anharmonic environment. We show that a strong anharmonicity produces two different effects depending on the environment temperature: for high temperatures, the system suffers a strong dephasing, while for low temperatures there is a strong information back-flow (as measured by the Breuer-Laine-Piilo (BLP) non-Markovianity measure). Both dephasing and back-flow are particularly significant when the anharmonic potential allows environment states very close to the  dissociation limit. In contrast, the information back-flow is suppressed when assuming the environment to be Gaussian. In this regard, we find that the Gaussian approximation is particularly poor at low temperatures. 
\end{abstract}
\maketitle

\section{Introduction}
An accurate description of hybrid systems such as molecular ensembles containing different types of degrees of freedom is still challenging. An often convenient approach is to consider some of these freedoms as an \textit{open} system, while others that operate at a different time scale are treated as an environment and described by its statistical properties \cite{breuerbook,brandes2005,rivas2011a,rivas2014,breuer2015,devega2015c}. The open system may represent for instance an electronic component while the environment describes a set of nuclei. In other situations the system is an impurity coupled to a complex environment that can be either in condensed matter form, a liquid or a gas. One of the most successful approaches consists in approximating such environment as a set of harmonic oscillators, an idea that was first put forward by Feynman and Vernon \cite{feynman1963a}. In this context, the best known model is the spin-boson \cite{leggett1987}, which is very extended to describe energy transfer between two or more molecules (or molecular states) in the presence of an environment \cite{gilmore2006,nazir2009}. Examples of these systems are antenna molecules within photosynthetic complexes that are coupled to a protein environment \cite{xu1994,plenio2008,ishizaki2009,rebentrost2009,devega2011a,mohseni2013}, or electron-transfer reactions between electronic donor and acceptor states that are conditioned by the motion of nuclear degrees of freedom \cite{tang1993,barbara1996,mavros2016}. In both cases, the protein and nuclear degrees of freedoms are represented by a harmonic bath producing dephasing. Other situations where electron transfer is affected by an environment include transport in polymers \cite{bredas2004}, dynamics of organic molecules within solar cells \cite{zhao2012}, or impurities in strongly condensed matter or liquids \cite{tanimura1993}.

However, the emergence of new scenarios and the rapid development of experimental techniques such as time-resolved nonlinear spectroscopy are giving more information on the dynamics of complex molecular systems, therefore requiring more detailed models and analysis \cite{okumura1996}. For instance, when the electronic dynamics occur in the presence of nonpolar liquids or low-frequency intramolecular modes (such as torsional motion) the harmonic approximation is known to fail \cite{tang1993,evans2000,lockwood2002,wang2004,wang2007,thoss2006}. Also, nuclear environments may have a complex structure where some modes are highly anharmonic, while the remaining ones can be treated as harmonic and considered to be linearly coupled to the anharmonic ones \cite{garg1985,hsieh2014}. Another relevant type of environment that is present in condensed matter physics and quantum technological devices corresponds to a set of spins or spin bath \cite{prokofev2000}. For these spin environments it is well known that the statistics (i.e. the behaviour of different order correlation functions of the environment coupling operators), is highly non-Gaussian, and therefore their description in terms of a spectral density is highly inaccurate. In such context of quantum technological devices, where noise-optimized quantum control is required, an accurate characterization of the environment and its non-Gaussian features is essential \cite{norris2016,sung2019}.
\begin{figure}[ht]
\includegraphics[width=0.9\linewidth]{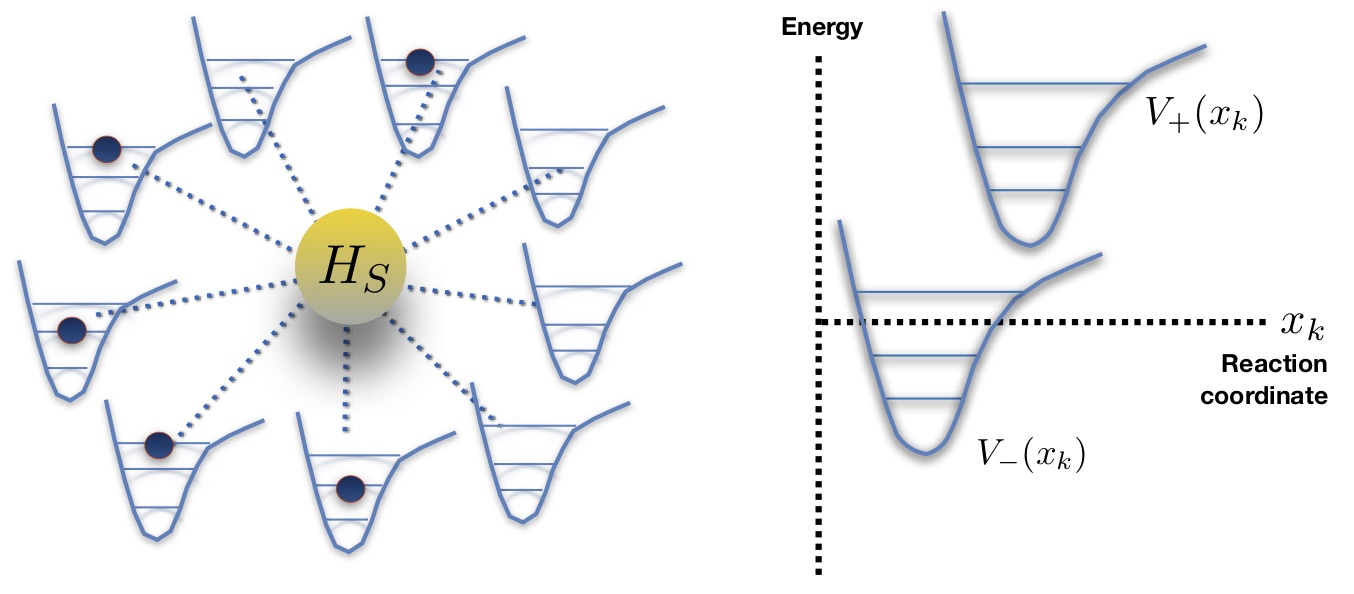}
\caption{(Color online) Left panel represents a scheme of the open quantum system coupled to an environment of anharmonic oscillators, which we model with a Morse potential. Right panel shows a schematic of the upper and lower level energy potential surfaces in one of the multi-dimensional reaction coordinates $x_k$ that describes an energy-transfer reaction between the electronic levels $|+\rangle$ and $|-\rangle$.   \label{morse}}
\end{figure}

In this paper, we consider an impurity coupled to an environment consisting of a set of independent Morse oscillators \cite{lopez2011,hsieh2018} producing dephasing. Advantageously, the Morse oscillator Hamiltonian is analytically solvable \cite{morse1929}, giving rise to a spectrum of discrete energy levels, or bound states, and a set of scattering states. Such scattering states form a continuum that produces an additional decoherence and is describable as a harmonic bath \cite{burkey1984,kazansky1997}. Here, we are not interested in a complete description of the whole Morse spectrum, but rather we focus in its discrete part. This allows us to make a systematic study of the effects of the anharmonicity on the open system dynamics: by tuning the anharmonicity of the potential we will be able to vary from a regime where each environment oscillator has a large number of equally spaced bound states, thus a harmonic limit, to a highly anharmonic situation with only two bound states. In other words, our model allows us to describe the dephasing dynamics of an impurity coupled to a harmonic bath and to a spin bath, as well as all the intermediate anharmonic regimes. As shown in Fig. (\ref{morse}) the Hamiltonian describes a star configuration, with the impurity coupled to a set of independent oscillators, such that the harmonic and spin limits correspond to the pure dephasing version of the spin-boson and central spin models, respectively. 

The paper is organized as follows. We first discuss general concepts of the model in Sec. \ref{general} as well as the dynamical map that describes the reduced dynamics of the impurity. Then we divide the analysis in three parts: In Sec. \ref{sec_dephasing} we discuss the changes in the impurity dephasing time, that is intimately related to the persistence of quantum mechanical properties such as coherence and entanglement along the evolution, when considering two impurities initially entangled \cite{yu1992}. In Sec. \ref{sec_flow} we consider the ratio between the flow of information inwards and outwards of the impurity \cite{cosco2018}, which determines the amount of information that is lost from the system into the environment. In turn, such ratio is strongly linked to the presence of BLP non-Markovianity  \cite{rivas2014,breuer2015,li2018}, a concept that has gained increasing attention in various contexts including quantum information and quantum metrology, biological systems, ultra-cold atoms or quantum thermodynamics \cite{devega2015c}. In Sec. \ref{sec_gaussian} we analyze the degree in which the impurity dynamics can be well described with a Gaussian map, such that it only depends on the environment second order moment or correlation function. This will allow us to explore when will the non-Gaussianity of the environment become more or less relevant depending on the environment temperature and the anharmonicity parameter. 
Finally, in Sec. \ref{conclusions} we draw some conclusions and outlook.

\section{A spin coupled to a Morse oscillator environment}
\label{general}
We consider a system with a total Hamiltonian $H=H_S+H_E+H_I$. This corresponds to an impurity with Hamiltonian ($\hbar=1$)
\bea
H_S=\w_s \s_z,
\label{hs}
\eea
where $\s_z$ is a spin ladder operator that can be written in terms of the impurity internal basis $|\pm\rangle$ as $\sigma_z=|+\rangle\langle+|-|-\rangle\langle -|$, 
coupled to an anharmonic environment composed of independent Harmonic oscillators with a  Hamiltonian $H_E=\sum_k H_k$ ($k \in \{1,\dots, 40\}$), where 
\bea
H_k=\frac{p_k^2}{2m} + D_k \left(e^{-2\alpha_k x_k}- 2 e^{-\alpha_k x_k}\right), 
\label{HK}
\eea
is describing each $k$-th oscillator in terms of a Morse potential. Here we have defined the position operator as $x_k=\sqrt{\alpha_k/2m\omega_k}(b_k+b_k^\dagger)$, where $b_k$ ($b_k^\dagger$) the standard harmonic annihiliation (creation) operators. The depth and the width of the Morse potential is determined by two different parameters, $D_k$ and $\alpha_k$, respectively. Notice that this represents a specific case of the right panel of Fig.\ref*{morse} in which the potential part $V(x_k)=D_k \left(e^{-2\alpha_k x_k}- 2 e^{-\alpha_k x_k}\right)$ is centered at the origin and equal for both impurity levels. Finally, the coupling is described through the Hamiltonian 
\bea
H_I=S \otimes B, 
\eea
where $S:=\sigma_z$ and $B$ are system and environment coupling operators, respectively. Just like for the free Hamiltonian we can write $B=\sum_k B_k$, with  
\bea
B_k&=&g_k(b_k+b_k^\dagger),
\label{BK}
\eea
with $g_k$ a constant that determines the coupling strength of the $k$-th oscillator to the system. 
Thus, we consider the coupling as linear and therefore proportional to the position operator (also called reaction coordinate in the context of electron transfer) of each Morse oscillator $B_k\sim x_k$ \cite{okumura1996,wang2004,thoss2006,okumura1996,wang2007,mavros2016,kananenka2018}.

\subsection{Properties of the Morse potential}

Proposed in 1929 \cite{morse1929} the Morse potential was one of the first empirical models to describe anharmonicities in the vibration of diatomic molecules, and is now widely used to describe complex anharmonic effects \cite{lopez2011,kananenka2018,hsieh2018}. The Morse potential presents various interesting features. First of all it is analytically solvable, yielding a finite number of bound states with negative energies given by 
\bea
E_{kn}:= -\frac{\w_k \Lambda}{2} - \frac{\w_k}{2\Lambda}\left(n+\frac{1}{2}\right)^2 + \w_k \left(n+\frac{1}{2}\right). \label{eq:5}
\eea
These energies depend on the depth and the width of the potential through a new parameter $\Lambda$, such that $D_k= \frac{\w_k \Lambda}{2}$ and $\alpha_k= \sqrt{\frac{\w_k}{\Lambda}}$. 
Thus, $\w_k$ is the frequency of the harmonic part of the potential, while the parameter $\Lambda$ tunes the anharmonic component. Note that while we consider the same anharmonicity parameter for all environment oscillators, their frequencies are linearly distributed.

Besides the bound states, the Morse potential present a continuous spectrum of positive energy eigenstates, also known as scattering states, that will not be considered here for simplicity. Overall, they will correspond to a continuous bath of free particles, to which the impurity may be coupled too and which may produce additional dephasing. In the new eigenstate basis, with the $n$-th energy eigenstate of the $k$-th oscillator written as $\ket{kn}$ and the respective eigenenergies $E_{kn}$, we find that Eqs. (\ref{HK}) and (\ref{BK}) can be written as 
	\bea
	H_k&=& \sum_{n} E_{kn} \ket{kn} \bra{kn}, \\
	B_k&=& \sum_{n,m} c^k_{nm} \ket{kn} \bra{km}, 
	\eea
where $c^k_{nm}=\langle k n|B_k|km\rangle$ are the coefficients for the bath part of the interaction operator.

\begin{figure}[ht]
\includegraphics[width=0.9\linewidth]{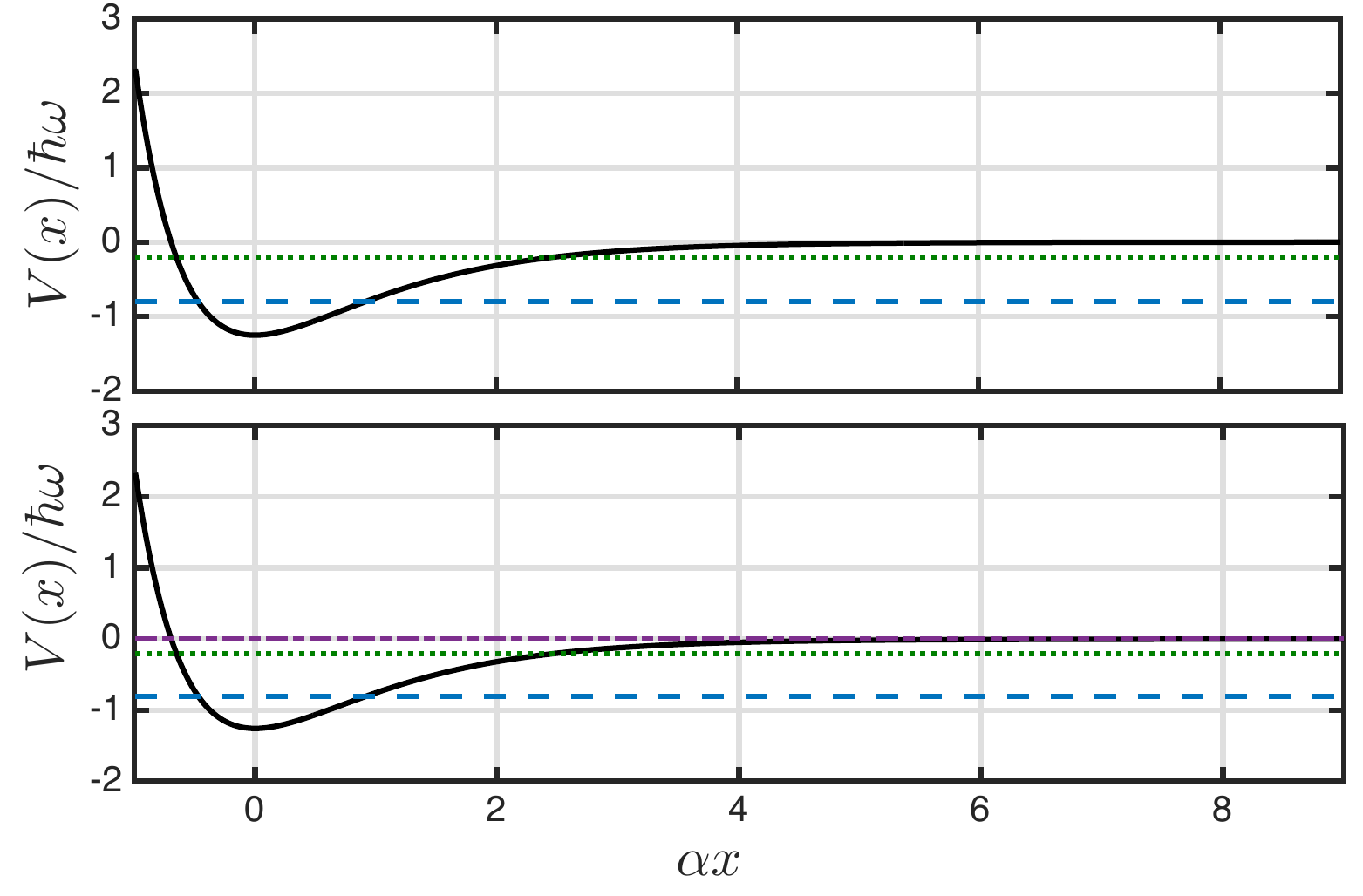}
\caption{(Color online) Morse potential and eigenvalues of the bound states for different values of the anharmonic parameter: Region I: $\Lambda\equiv\Lambda^{n=2,\epsilon=0}=2.5$ (upper panel) displaying two bound states far from the dissociation limit, and region II, $\Lambda\equiv \Lambda^{n=2,\epsilon=0.1}=2.51$ (lower panel) giving rise to the same bound states and a third one that corresponds to a weakly bounded state, close to the dissociation limit. \label{thesystem}}
\end{figure}

A second interesting property is that the number of bound states is uniquely given  by the integer part of $\Lambda +\frac{1}{2}$, when $\Lambda + \frac{1}{2}$ is not an integer. When $\Lambda+\frac{1}{2}$ is an integer, the number of bound states is $\Lambda-\frac{1}{2}$. This allows us to, by simply varying $\Lambda$,  tune our bath between a spin bath limit, corresponding to $\Lambda \in ]3/2,5/2]$, to a harmonic limit, recovered for $\Lambda \rightarrow \infty$. Note that at 3/2 the Morse oscillator has only one bound state.
Between these limits, we also find two different types of regions in $\Lambda$ which will determine very strongly the nature of the dynamics:
\begin{enumerate}[label=\Roman*)]
\item The regions where all environment states are strongly bounded, corresponding to $\Lambda\equiv \Lambda^{n,-\epsilon}=n+1/2-\epsilon$, with $n$ integer and $0\le|\epsilon|\ll 1$ a real number. 
\item The regions where a new bound state is formed, which moreover is weakly bounded, corresponding to $\Lambda\equiv \Lambda^{n,\epsilon}=n+1/2+\epsilon$, with $n$ integer and $0< \epsilon\ll 1$ a positive real number. 
\end{enumerate}
A typical situation in both regions can be observed in Fig. \ref{thesystem} for the case of $n=2$, when choosing $\epsilon=0$ (region (I)) and $\epsilon=0.1$ (region (II)).

\subsection{Properties of the open system dynamics}

We now consider an initially decorrelated state so that the total density operator is written as $\rho(t=0)= \rho_0 \otimes \rho_E^\beta$, where $\rho_0$ is the initial state of the system, and the environment is in a thermal equilibrium $\rho_E^\beta :=  e^{-\beta H_E}/\Ttr_E\{e^{-\beta H_E}\}$, with the inverse temperature $\beta=1/k_B T$ and $k_B$ is the Boltzmann constant.
In the general case, the reduced density matrix of the system in the interaction picture can be written as 
\bea
\rho_s(t)=\Ttr_E\{{\mathcal U}^{-1}(t,t_0)\rho_s(0)\otimes\rho^\beta_B(0){\mathcal U}(t,t_0)\},
\label{red}
\eea
where ${\mathcal U}(t,t_0)$ is the evolution operator in interaction picture, which can be expanded with a Dyson series as
\bea
{\mathcal U}(t,t_0)&=&1-i\int_{t_0}^tdt_1 H_I(t_1)\cr
&+&(-i)^2\int_{t_0}^tdt_1\int_{t_0}^{t_1}dt_2 H_I(t_1)H_I(t_2)+\cdots.
\eea
Considering this expansion as well as the case $H_I=SB$, we find that the time evolution of $\rho_s(t)$ can be re-written in terms of an infinite series of $l$-th order correlation functions which at zero times have the general form
\bea
C^l(0)=\Ttr_E\{B^l \rho_E\} = \sum_{k_1,\dots, k_l} \Ttr_E\{B_{k_1} \cdots B_{k_l} \rho_E\}.
\label{correlation_gen}
\eea
Here, the sum covers all configurations of $\{k_1,\dots, k_l\}$, and can be decomposed into sums over sectors where different numbers of those $k$ coincide. As noted above, our model describes a linear system-environment coupling, such that for each oscillator $B_k\sim x_k$, with $x_k$ the position operator. Three important consequences can be extracted from such linear coupling to an anharmonic environment:
\begin{enumerate}[label=(\roman*)]
\item The first order term ($l=1$) in the expansion (\ref{correlation_gen}) needed to recover Eq. (\ref{red}) is non-zero. This is because the coupling is linear and the environment Hamiltonian describes an asymmetric potential, such that $\left\langle B \right\rangle_E^\beta := \ttr_B\{B\rho_E^\beta\}\neq 0$. This fact can be remedied by considering a renormalized version of the Hamiltonian, such that $H=\tilde{H}_S+H_E+\tilde{B}\sigma_z$, where $\tilde{H_S}=\omega_s\sigma_z+\sigma_z\left\langle B \right\rangle_E^\beta$, and $\tilde{B}=B-\left\langle B \right\rangle_E^\beta$. Hence, the correlations involved in Eq. (\ref{red}) are equal to the ones in Eq. (\ref{correlation_gen}) but replacing each $B_k$ by their renormalized counterpart $\tilde{B}_k$. 
\item After the first order term, the most important contribution, i.e. the second order term or correlation function
\bea
\alpha(t-s)=\Ttr_E\{\tilde{B}(t)\tilde{B}(s)\rho_E^\beta\},
\label{correlation}
\eea
has a real part that does not decay to zero at finite temperatures but to a constant positive value.
In terms of the eigen-values and eigenvectors of the environment $\{E_{kn},|kn\rangle\}$, the correlation function (\ref{correlation}) can be formally written as $\alpha(t)=C(t)+C^0$, where 
			\bea
			C^0&=&\sum_{k,n} \label{4.1.26} |\bra{kn}\tilde{B}_{k}\ket{kn}|^2\frac{e^{-\beta  E_{kn}}}{Z_k} ,
			\label{offset}
			\eea
			corresponds to an offset, while 
			\bea
			C(t)&=&\sum_k \sum_{p\neq n} |\bra{kn}\tilde{B}_{k}\ket{kp}|^2\frac{e^{-\beta  E_{kn}}}{Z_k} e^{i\Delta_{np}^kt}, \label{timedep}
			\eea
			corresponds to a time-dependent component, where we have defined $\Delta_{np}^k=E_{kn}-E_{kp}$. When we have a sufficiently large number of interfering phases, such time dependent factor decays. However, the time independent contribution will produce a real offset that will only be cancelled in the zero temperature limit, i.e. when we have that $C^0=\sum_{k} |\bra{k0}\tilde{B}_{k}\ket{k0}|^2=|\langle \tilde{B}\rangle^{\infty}_E|^2=0$.
\item The offset is particularly important in regions of type (II), i.e. for values of $\Lambda$ for which a weakly bound state exists, since these correspond to highly asymmetric bound states leading to a large overlap $\bra{km}\tilde{B}_{k}\ket{km}$, where we define $m$ as the index of such weakly bound state. The presence of the weakly bounded state has two important consequences: First, it holds that:
\bea
\lim_{\varepsilon\rightarrow 0^+} \bra{m}x\ket{m}&=&  \ln(2m+1) -\lim_{\varepsilon\rightarrow 0^+}\psi(2\varepsilon) \cr
&-&\psi(1) + \psi(m+1)=\infty,
\label{diag}
\eea
where we have just inserted the matrix element given in (\ref{eq:3.1}) and took the limit. In consequence, the offset becomes very large nearby such limit, and when the temperature is high enough so as to have a large initial population in the upper energy level. The second consequence is that all the transitions to the bound state with highest energy are suppressed. To show this, we consider a bound state $|l\rangle$ having lower energy than $|m\rangle$, then 
\bea
&\lim_{\varepsilon\rightarrow0^+} &\bra{l}x\ket{m}=
\frac{2(-1)^{m-l+1}}{(m-l)(m-l)}\cr
&\times&\sqrt{\frac{m!(m-l)\Gamma(m+1)}{l! \Gamma(2m-l+1)}} \lim_{\varepsilon\rightarrow0^+} \sqrt{\varepsilon}
=0.
\label{transition}
\eea
Here we took (\ref{eq:3.2}), inserted $N=m+\varepsilon$ and used that the functions containing $\varepsilon$ are continous functions so we could pull the limit into them. The consequence is that the time-dependent term of the correlation function, given by Eq. (\ref{timedep}) is not affected by the presence of the bound state, contrary to the offset part.

\item So far we have focused on the first and the second order moments of the expansion (\ref{correlation_gen}). However, in the present anharmonic case the structure of such expansion is much richer than in the harmonic case. First, the terms in $\Ttr_E\{\tilde{B}_{k_1} \cdots \tilde{B}_{k_l} \rho_E\}$  having an odd number of identical $k_j$ do no longer vanish. That is, $\Ttr_E\{\tilde{B}_{k}^{2n+1} \cdots \tilde{B}_{k} \rho_E\}\neq 0$, when $n$ is integer. Moreover, the even order components can no longer be decomposed as products of second order components of the form (\ref{correlation}). Naturally, these two properties are fulfilled in the harmonic bath case, leading to system dynamical equations that can be written solely in terms of the correlation function (\ref{correlation}). Indeed, Wick's theorem holds in the harmonic case, since there $H_k$ is quadratic and one can decompose $B_k(t)=e^{iH_k t}B_k e^{-iH_kt}=f_k(t)b_k(0)+f^*_k(t)b^\dagger_k(0)$ (with $f_k(t)$ a time-dependent function). Away from this limit, such a decomposition is no longer possible and Wick's theorem can no longer be applied. 
\end{enumerate}

One way to diminish the relative weight of higher order terms with respect to the second order one (\ref{correlation}) is to choose  an appropriate scaling for the coupling strengths $g_k$ present in the environment coupling operators $B_k$. In detail, if we choose them to scale as $g_k\sim1/\sqrt{K}$, where $K$ is the number of oscillators, it can be shown that higher order correlations will eventually vanish in the limit of large $K$, an idea that was originally proposed in \cite{makri1999a} and that is further discussed in \cite{bramberger2019b}. 

In our case, we consider a fixed number of oscillators $K=40$ and explore to which extent the higher order terms are relevant to the description of the system. In order to be consistent with the harmonic oscillator limit, we further consider the standard choice in this limit for the frequency distribution and coupling strengths, 
\bea
\w_k &:=& \frac{2\w_c}{K} k, \\
g_k &:=& \sqrt{\frac{2\w_c}{K} J(\w_k)}.
\eea
Here $J(\w)$ is the spectral density, which we consider to be of ohmic type 
\bea
J(\w) := \Theta\left(2\w_c-\w\right) \eta \frac{\w}{\w_c} e^{-\frac{\w}{\w_c}},
\label{spectral}
\eea
with a special hard cut and $\w_c$ is a cut-off frequency, and $\eta$ is a parameter with which we modulate the strength of the system-environment coupling. For the harmonic case, the chosen linear discretisation $\omega_k=k\Delta\omega $, with $\Delta\omega=2\omega_c/K$, gives rise to a revival time (which is the time at which finite size effects of the environment will start to occur) that in our case is $T=\pi/\Delta\omega\approx 20$.

\subsection{The dynamical map}

Because we are considering pure dephasing, we have that $\left[H_S,S \right]=0$, which implies that the  Hamiltonian is in block diagonal form and can be written as 
\bea
H&=& P_{+} \bigg(\w_s + \sum_k H_k^+\bigg) +P_{-} \bigg(-\w_s+\sum_k H_k^-\bigg)\label{eq:projectors}
\eea
%
%
where we have defined the projectors $P_\pm:=\ket{\pm}\bra{\pm}$ in terms of the eigenstates $\ket{\pm}$ of $\s_z$ with eigenvalues $\pm1$, and we have decomposed both $H_E$ and $B$ in terms of local operators, $H_k^\pm:=H_k \pm B_k$ with $H_k$ corresponding to the $k$-th oscillator. Thus, the time-evolution operator can be computed as 
\bea
e^{iHt}&=&
P_- \; e^{-i\w_st} \prod_k e^{iH_k^- t} + P_+ \; e^{i\w_s t} \prod_k e^{iH_k^+ t}.
\label{evol}
\eea
To obtain this expression, we have used the fact that the two terms in (\ref{eq:projectors}) commute, as well as the locality of the terms in the exponential to factorize it. Thus, computing Eq. (\ref{evol}) requires only having to exponentiate local operators which can be done very efficiently. Indeed, the total Hilbert space dimension scales like $d^K$, where $d$ is the local dimension, and we only have to exponentiate matrices of size $d$. Thus, the total density operator $\rho(t)=e^{-iHt} \rho_0 \otimes \rho_E^\beta  e^{iHt}$ can be rewritten as 
\bea
\rho(t)&=& P_+ \rho_0 P_+ \otimes e^{-iH^+t} \rho_E^\beta e^{iH^+t} \nonumber \\
&+& P_- \rho_0 P_- \otimes e^{-iH^-t} \rho_E^\beta e^{iH^-t} \nonumber \\
&+& P_- \rho_0 P_+ \otimes e^{-iH^-t} \rho_E^\beta e^{iH^+t} e^{2i\omega_s t} \nonumber \\
&+& P_+ \rho_0 P_- \otimes e^{-i H^+t}\rho_E^\beta e^{iH^-t} e^{-2i \omega_s t},
\eea
where we have defined $H^\pm:= \sum_k H_k^\pm$. Taking the partial trace over the bath degrees of freedom yields the reduced density matrix of the system, which has the following form
\bea
 \rho_s(t)=\Phi(t)\left[\rho_s(0)\right]=\left( \begin{array}{cc}
\rho_s^{11}(0) & \chi(t) \rho_s^{12}(0) \\[5pt]
\chi^*(t) \rho_s^{21}(0) & \rho_s^{22}(0) \end{array} \right)
\label{rhos}
\eea
where we have defined the decaying factor as 
\bea
\chi(t)=e^{2i \omega_s t} \prod_k \ttr_k\left(e^{-iH^-_k t}\rho_k^\beta e^{iH^+_k t}\right),
\label{decay} 
\eea
where $\rho_k^\beta$ is the thermal state of the $k$-th oscillator with respect to its free Hamiltonian. We note that despite of the presence of an offset (\ref{offset}) which gives rise to an ill defined weak coupling master equation, we have numerically found that the map (\ref{decay}) is invertible for all the parameter regimes we considered here (not shown), which suggests that a time-local master equation is still well-defined \cite{hall2014}.

\subsubsection{Harmonic and spin limits}

In the harmonic oscillator limit, the decay factor acquires the usual form 
\bea
\chi(t)=e^{2i\omega_s t-\Gamma(t)}, 
\label{chiharm}
\eea
where  
\bea
\Gamma(t) &:=&8 \sum_k  \frac{g_k^2}{\omega_k^2}\sin^2\left(\frac{\w_k t}{2}\right)\coth\left(\frac{\beta \omega_k}{2}\right)\cr
&=&4\Rre\bigg\{\int_0^t ds\int_0^s du \alpha(s-u)\bigg\},
\label{gammaharm}
\eea
where $\alpha(t-s)=\Ttr_E\{\tilde{B}(t)\tilde{B}(s)\rho_E^\beta\}$. Moreover, note that in the harmonic limit $\tilde{B}=B$, since $\langle B\rangle^\beta=0$. In addition, the offset (\ref{offset}) of the correlation function vanishes.

In the opposite limit of a spin bath, i.e. when $1.5<\Lambda\leq2.5$, the operators $H_k$ and $B_k$ can be written in terms of the Pauli matrices $\sigma_j^k$ which are elements of the vector $\vec{\s}_k=(\unit_k,\sigma^k_x,\sigma^k_y,\sigma^k_z)$, where we have defined the first element $\s_0^k=\unit_k$ in terms of the unit operator in the Hilbert space of the $k$-th oscillator. In this representation, we find that 
\bea
H^+_k&=&\vec{c}_k\cdot\vec{\s}_k,\cr
H^-_k&=&\vec{d}_k\cdot\vec{\s}_k,
\eea
where each component of $\vec{c}_k$ and $\vec{d}_k$ is defined as $c^k_j = \frac{1}{2} \Ttr_k(H^+_k \s^k_j)$ and $d^k_j = \frac{1}{2} \Ttr_k(H^-_k \s^k_j)$, respectively. Thus, the  exponentials appearing in Eq. (\ref{decay}) can be simplified as 
\bea
e^{\pm i H^+_k t} &=& \bigg(\unit \cos\left( \left\Vert \vec{c}_k \right\Vert t\right) \pm i \frac{\vec{c}_k \cdot \vec{\s}_k}{\left\Vert \vec{c}_k \right\Vert} \, \sin\big( \left\Vert \vec{c}_k \right\Vert t \big)\bigg) e^{ \pm i c^k_0 t}, \cr
e^{\pm i H^-_k t} &=& \bigg(\unit  \cos\big( \Vert \vec{d}_k \Vert t\big) \pm i \frac{\vec{d}_k \cdot \vec{\s}_k}{\Vert \vec{d}_k \Vert} \, \sin\big( \Vert \vec{d}_k \Vert t \big)\bigg) e^{\pm i d^k_0 t}\cr
&&\label{eq:spindeph2}
\eea


\section{Dephasing time}
\label{sec_dephasing}
We first analyze the dephasing time of the system. i.e. the decaying of the off-diagonal elements of the reduced density matrix. For all the numerical results we choose frequency units of $\w_s=1$. Moreover, we consider the initial state 
\bea
\rho_0=\frac{1}{2} \left(\unit + \frac{1}{2} \s_x\right).\label{eq:5init}
\eea
Although this choice is rather arbitrary, it ensures that there are initial coherences that allow us to analyze dephasing. In other words our analysis is independent from the initial condition as long as it contains coherences. We define the decay time $\tau_d$ as the time such that 
\bea
\frac{\lvert\bra{0}\rho_S(\tau_d)\ket{1}\rvert}{\lvert\bra{0}\rho_S(t=0)\ket{1}\rvert}=\frac{1}{10},
\eea
where the reduced density matrix is given by Eq. (\ref{rhos}). 
\begin{figure}[ht]
\includegraphics[width=1.\linewidth]{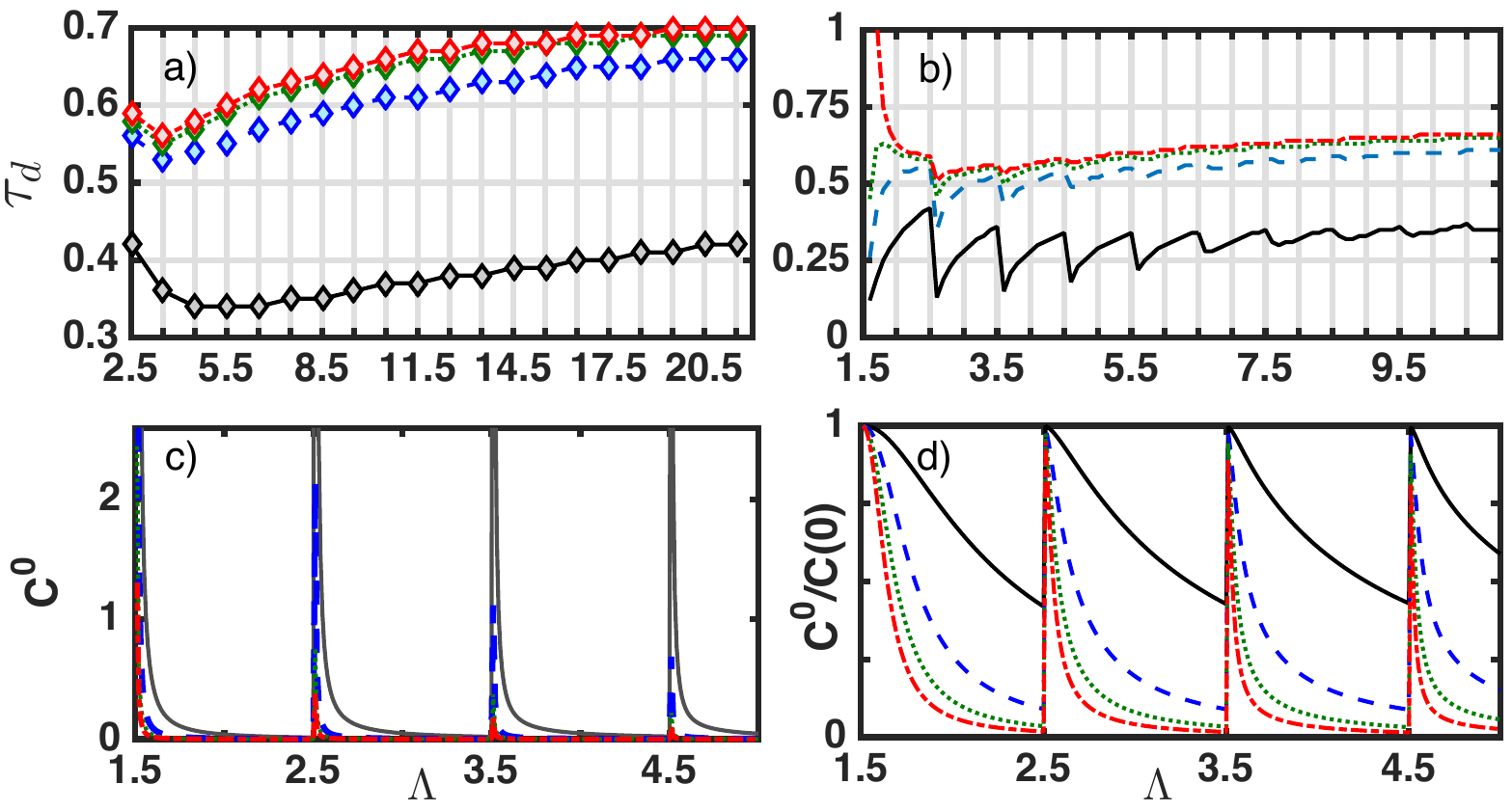}
\caption{(Color online) Upper panels represent the dephasing rate for a coupling strength of $\eta=2$ in Eq. (\ref{spectral}). The the left panel has low resolution with values of $\Lambda$ separated by a step $\delta \Lambda=1$, while the right has a high resolution with $\delta \Lambda = 0.1$. The lower panels  represent the offset (left) and relative value of the offset $C^0/C(t=0)$ (right), for high resolution $\delta \Lambda=0.01$ and $\eta=0.01$. All panels display the curves for different inverse temperatures, $\beta=1,4,7,10$ corresponding to solid black, dashed blue, dotted green, and dot-dashed red, respectively. \label{dephasing}}
\end{figure}

Figs. \ref{dephasing}(a,b) display the decay time $\tau_c$ for different values of the anharmonicity parameter $\Lambda$ and different initial temperatures of the bath. For the following discussion it is important to remember that as $\Lambda$ increases, the anharmonicity decreases.
In detail, in Fig. \ref{dephasing}(a), we explore the values of $\Lambda\equiv\Lambda^{n,\epsilon=0}=n+1/2$, which correspond to type (I) regions where the highest energy bound state is as far as possible from the dissociation limit, i.e. as strongly bounded as possible. With this choice, we find that $\tau_c$ has a quite smooth behavior with the anharmonicity, displaying a general tendency to increase towards the harmonic limit. Such trend reverses towards higher anharmonicities, where the number of bound states becomes increasingly limited and the dephasing slows down again.

As shown in Fig. \ref{dephasing}(b), such a smooth behavior is disrupted when including a finer grid of $\Lambda$ values, now taking closer points spaced by $\delta\Lambda=0.1$.  One can clearly identify that the regions of type (I) end up with a peak at values $\Lambda=n+1/2$ (where the top of the peaks correspond to the points shown in Fig. \ref{dephasing}(a)), followed by a dip at values $\Lambda^{n,\epsilon}$ with $\epsilon$ very small, i.e. when the regions (II) begin. Indeed, as discussed in Fig. \ref{dephasing}(a), at high anharmonicities each oscillator has only a few energy levels, which in principle hinders the dephasing. However, when entering each region (II) a new bound state is formed, and the offset becomes so large that it compensates for this effect and leads to a strong dephasing. 

Another interesting feature of the figure is that in the spin region the decay time goes to infinity at low temperatures and at low $\varepsilon$ at the beginning of region (II). Indeed, we have seen in Eq. (\ref{transition}) that all the transitions to the bound state with the highest energy are suppressed in regions (II). Thus, when we only have two bound states the limit $\varepsilon\rightarrow0$ gives rise to bath operators $H^\pm_k$ that are diagonal. To see this, we find that for $\Lambda=1.5+\varepsilon$ and small $\varepsilon$,
\bea
\lim_{\varepsilon\rightarrow0^+} c^k_x &=& \lim_{\varepsilon\rightarrow0^+}\left( \ttr_k(H^+_k \s_x) - \ttr_k(B_k \s_x)\right) =0, \cr
\lim_{\varepsilon\rightarrow0^+} c^k_y &=& \lim_{\varepsilon\rightarrow0^+}\left( \ttr_k(H^+_k \s_y) - \ttr_k(B_k \s_y)\right) = 0 ,
\eea
and similarly $\lim_{\varepsilon\rightarrow0^+} d^k_x=0$ and  $\lim_{\varepsilon\rightarrow0^+} d^k_y=0$. In addition, 
\bea
c^k_z &=&  \ttr_k(H^+_k \s_z) + \ttr_k(B_k \s_z), \cr
d^k_z &=&  \ttr_k(H^-_k \s_z) - \ttr_k(B_k \s_z),  \label{eq:limdephcoeff}
\eea
and their limit $\varepsilon\rightarrow0^+$ will diverge as predicted by the general case eq. (\ref{diag}). If we now consider the eqs. (\ref{eq:spindeph2}) for a very small $\varepsilon$, and ignore for this argument the phase that comes from the $c_0$ and the $d_0$ contribution because it does not affect the absolute value of the coherences, we find 
\bea
\lim_{\varepsilon\rightarrow0^+} e^{\pm i H^+_k t} &=& \unit_k  \cos\big(\phi_k^+ t\big) \pm i  \s_z \, \sin\big(\phi_k^+ t \big),\cr
\lim_{\varepsilon\rightarrow0^+} e^{\pm i H^-_k t} &=& \unit_k \cos\left(\phi^-_k t\right) \mp i \s_z \, \sin\left(\phi^-_k t \right),\label{eq:dephlim2}
\eea
where we have defined the phases $\phi_k^+=\lim_{\varepsilon\rightarrow0^+} \big\Vert \vec{c}_k \big\Vert$ and $\phi^-_k=\lim_{\varepsilon\rightarrow0^+} \big\Vert \vec{d}_k \big\Vert$. Strictly speaking the above equations are ill-defined in the limit $\lim_{\e\rightarrow 0}$. For this  reason, we look at small but not infinitesimal $\epsilon$. 
Also, we shall note that the sign between the two terms in the second equation is flipped as compared to the first. This is because the $\ttr_k(H_k\s_z)$ term in equation (\ref{eq:limdephcoeff}) becomes insignificant compared to the $\ttr_k(B_k \s_z)$ term. The phases $\phi_k^\pm$ are going to infinity towards the limit $\varepsilon\rightarrow0^+$. However, if we stay at a very small, yet not tiny value of $\varepsilon$ we can consider that $\phi_k^\pm\approx \Gamma^\pm$, where $\Gamma^\pm$ are two large phases, and define $\Gamma_T=\Gamma^-+\Gamma^+$, such that in Eq. (\ref{decay}) we now have 
\bea
\ttr_k\left(e^{- i H^-_k t}\rho_k^\beta e^{ i H^+_k t} \right)&&\approx \cos\left(\Gamma_T t \right)
+ i \sin\left(\Gamma_T t\right) \ttr_k(\s_z \rho_k),\cr
&&
\eea
where $\ttr_k(\s_z \rho_k)= \tanh\left(\frac{\beta \Delta E_k}{2}\right)$, with $\Delta E_k$ the energy difference of the $k$-th spin. Thus at zero temperature we simply find
\bea
\left\vert\ttr_k\left(e^{- i H^-_k t}\rho_k^\beta e^{ i H^+_k t} \right)\right\vert&\approx& \cos^2\left(\Gamma_T t \right) + \sin^2\left(\Gamma_T t\right) \cr
&=&1.
\eea
The reason why the dephasing time does not increase to infinity at the beginning of other type (II) regions in the plot, is because as soon as there are more bound states some off diagonal transitions to intermediate states are allowed, which gives rise to dynamics even in the zero temperature case.

The large offset at the beginning of regions (II) can be observed in Fig. \ref{dephasing}(c), and as argued before in point (iii) it is particularly relevant at high temperatures. From this plot, one may be tempted to believe that the offset is only important in such regions of type (II). However, when representing in Fig.(\ref{dephasing})(d) one can see that relative value of the offset with respect to the maximal value of the time dependent part, i.e. $C^0/C(t=0)$ is non-negligible in all regions of $\Lambda$, including the region (I) where there is no weakly bounded state. Thus, on the one hand the offset explains the short dephasing time at high temperatures ($\beta=1$) with respect to lower temperatures, and on the other hand, for all temperatures it explains the tendency for a shorter dephasing towards $\Lambda$ smaller. 

\section{Outflow and backflow of information}
\label{sec_flow}
\begin{figure}[ht]
\includegraphics[width=1.0\linewidth]{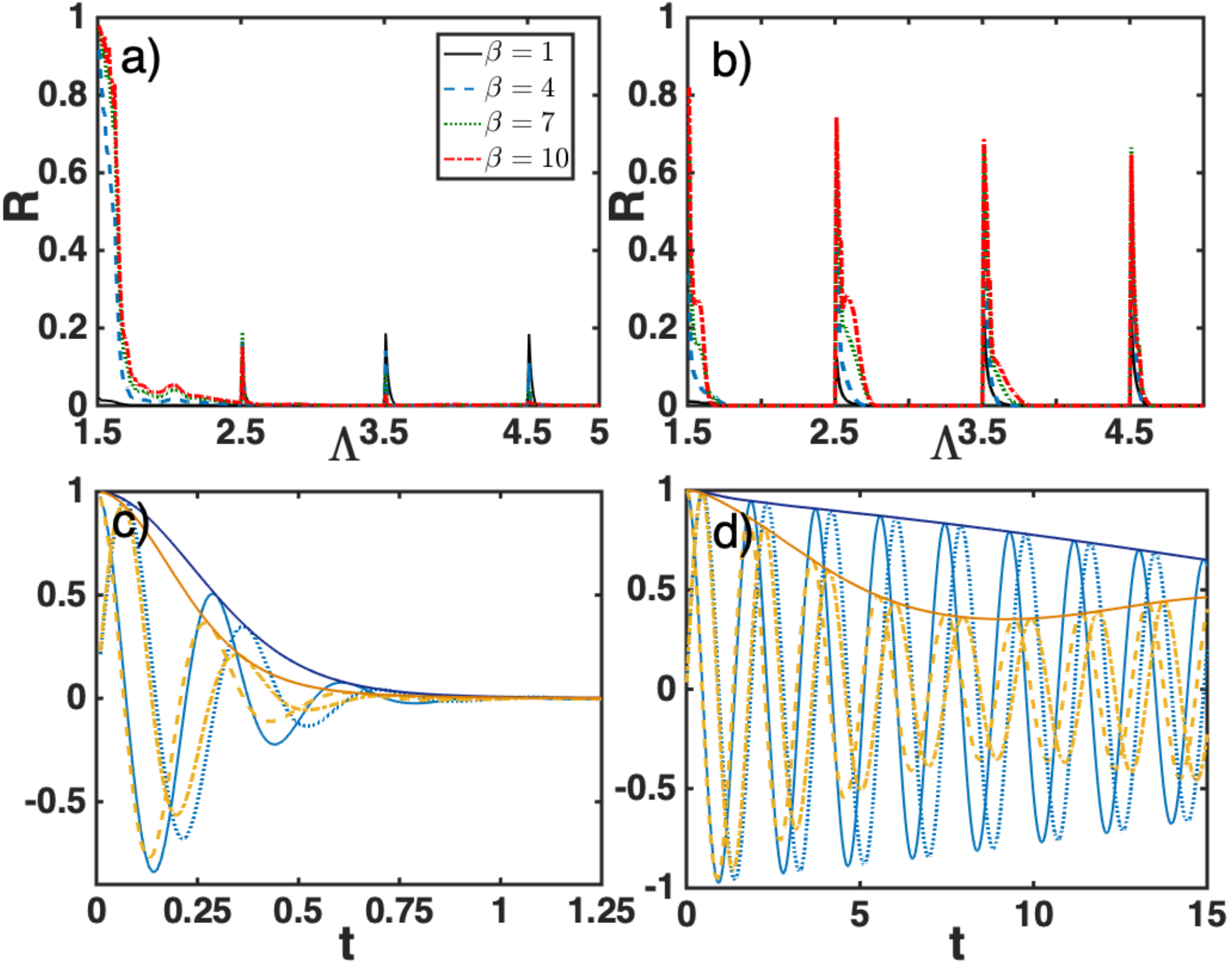}
\caption{(Color online) The upper panels represent the rate (\ref{rate}) for strong coupling $\eta=2$ (left upper panel) and weak coupling $\eta=0.01$ (right panel). The lower panels represent the evolution of $\chi(t)$ in Eq. (\ref{decay}) for $\beta=7$ and for strong coupling $\eta=2$ (left upper panel) and weak coupling $\eta=0.01$ (right panel). The oscillatory solid blue and dotted blue lines correspond to the real and imaginary parts of $\chi(t)$ for $\Lambda=2.5$, while the yellow dashed and dot-dashed lines correspond to the real and imaginary parts for $\Lambda=2.6$, respectively. Solid lines in the peaks represents the correspondings $|\chi(t)|$. 
\label{flow}}
\end{figure}

We now analyze how the outflow and backflow of information is affected by the degree of anharmonicity of the environment. To this aim, we consider that the back-flow of information, as obtained from the BLP measure can be written as \cite{cosco2018}
\bea
{\mathcal N}_{-}=\sum_n |\chi(t_{2n})|-|\chi(t_{1n})|,
\label{NM}
\eea
where $[t_{1n},t_{2n}]$ are the time intervals over which $|\chi(t))|$ increases, and $\chi(t)$ is given by Eq. (\ref{decay}). In a similar way as in \cite{cosco2018} we consider the ratio between the information backflow, given by Eq. (\ref{NM}) and the analogous quantity corresponding to the information flow to the environment, ${\mathcal N}_{+}$,
\bea
R={\mathcal N}_{-}/{\mathcal N}_{+}.
\label{rate}
\eea
The upper panel of Fig. (\ref{flow})(a) represents the ratio $R$ for different values of anharmonicity $\Lambda$ and for the same coupling value ($\eta=2$) considered in the dephasing time analysis. We find that the only region where there is a non-negligible back-flow of information is the highly anharmonic one, particularly the spin-bath region. Fig. (\ref{flow})(b) represents the same quantity but now for weaker coupling, $\eta=0.01$, showing that in this case the back-flow of information is also significant for higher values of $\Lambda$, particularly nearby the transition regions where a new environment bound state has been formed. 
Interestingly, the presence of back-flow is related to the following features:
\begin{itemize}
\item At high temperatures there is almost no back-flow. As seen before, the presence of a large offset gives rise to a fast dephasing time which in turn eliminates any possibility of back-flow and therefore non-Markovianity. 
\item At lower temperatures, the offset is not as important as to eliminate all the structure in the dephasing dynamics, and the maximal back-flow is observed precisely at the same values of $\Lambda$ where a new bound state has just been created (regions (II)). 
\end{itemize}

The dynamics of $\chi(t)$ can be observed in Fig. (\ref{flow})(c) (strong coupling) and Fig. (\ref{flow})(d) (weak coupling) for $\beta=7$. In detail, one can observe that while the real and imaginary parts of $\chi(t)$ are oscillatory, its absolute value is a monotonically decreasing function for $\Lambda=n+0.5=2.5$ and becomes non-monotonically decreasing (i.e. giving rise to back-flow) for $\Lambda=2.6$, i.e. for a Morse potential value having a weak bound state. 

An important comment is here in order. Since our environment is finite, a border effect is to be expected at a certain time. For $K=40$ oscillators, the revival time can be estimated to be $T=20$ in the harmonic limit (see \cite{devega2015b} and references therein for details), but it is much harder to compute in the anharmonic case. The reason is that, while in the harmonic case the revival time is the time at which the correlation function, having decayed, starts to grow again, in the anharmonic case we have seen that such function is not the only one that comes into play. For $\eta=2$ the system off-diagonal elements have already decayed to zero at all considered values, which allows to consider our analysis of dephasing time. However, this is not the case for smaller values, like $\eta=0.01$, where at $T=20$ there are curves that have not yet decayed. This means that the present analysis of forward and back-flow has been performed within a time-frame up to $T=20$ where neither we can consider that the system has always completely decayed, nor we can exclude border effects in the anharmonic limit. A rigorous analysis of such finite size effects produced by non-Harmonic environments is out of the scope here, but would be very interesting. In particular, it is interesting to ask whether the measured back-flow is due to such finite-size effects, or to the nature of the environment itself, as considered in the harmonic case.

Finally, we computed the non-Markovianity by considering the Gaussian map $\chi^{\textmd{Gauss}}(t)$ given by Eq. (\ref{chiharm}) and the correlation function (\ref{correlation}), finding that is zero for all parameter regimes considered. It remains to be further analyzed if Gaussian evolutions (or their related properties such as non-Markovianity) obey some extremality property like that of Gaussian states \cite{wolf2006}. 

\section{Non-Gaussian nature of the bath}
\label{sec_gaussian}
\begin{figure}[ht]
\includegraphics[width=1\linewidth]{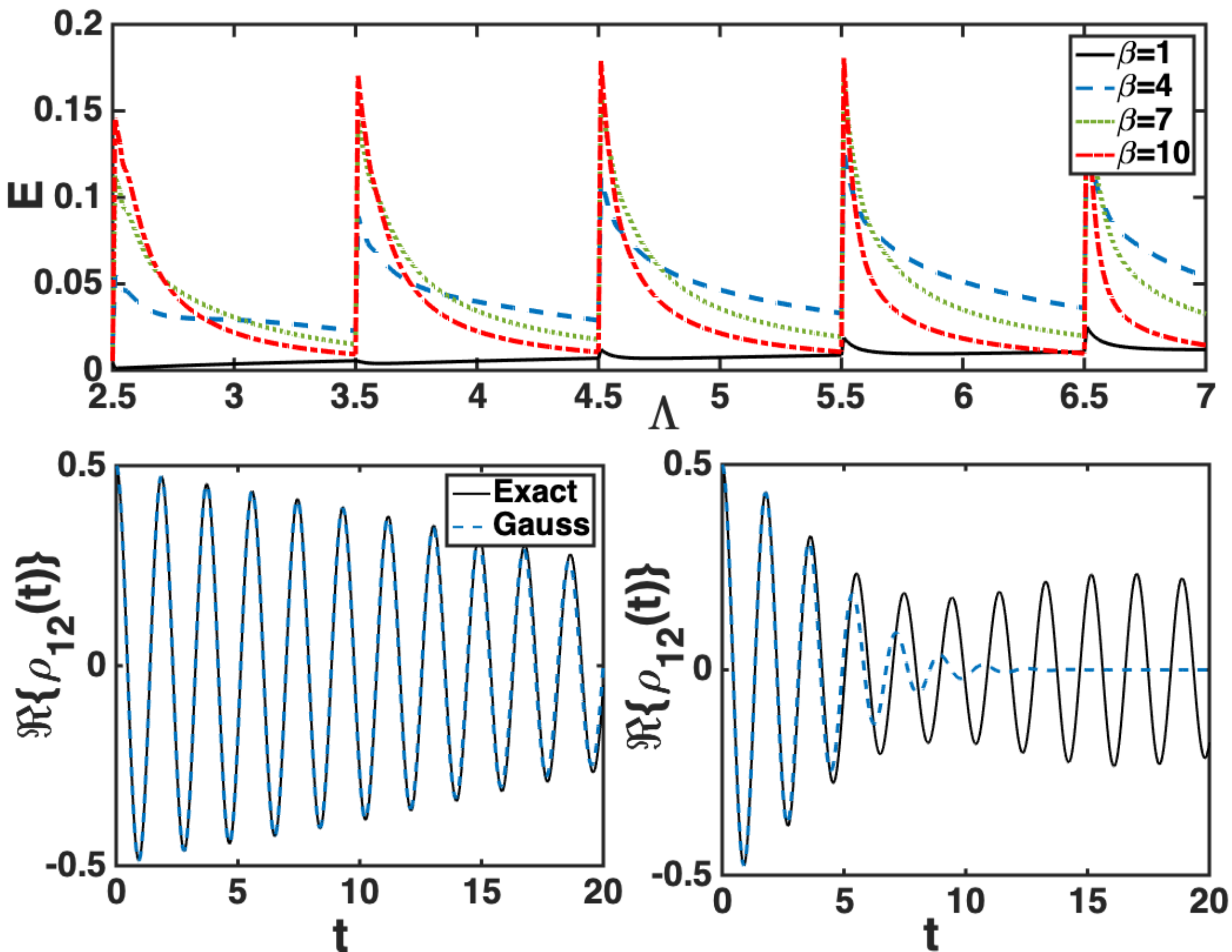}
\caption{(Color online) Time averaged error between the Gaussian evolution and the exact one, for $\eta=0.01$ and several temperatures, $\beta=1,4,7,10$ corresponding to solid black, dashed blue, dotted green and dot-dashed red, respectively. The maximal time taken is $T=20$, which corresponds to the time up to which, for the harmonic case, the system evolves without any border effect. The lower panels represent the evolution of the real part of the off diagonal elements of the density matrix as predicted by the exact map (solid black) and the Gaussian-approximated one (dashed blue). The left panel corresponds to $\Lambda=2.5$, and the right one $\Lambda=2.6$, both for $\beta=7$.\label{error_gaussian}}
\end{figure}

We have seen in the last sections that nearby the limit where a weakly bounded upper state is present in the environment oscillators there is a strong information back-flow, particularly at low temperatures. In this section we will further explore the nature of such non-Markovianity and show that it is linked to a strong non-Gaussianity of the bath. 
To this aim, we will compare the map (\ref{rhos}) computed with the exact non-Gaussian $\chi(t)$ corresponding to the Morse environment, Eq. (\ref{decay}), with the Gaussian $\chi^{\textmd{Gauss}}(t)$ (again as given by Eq. (\ref{chiharm}) with the correlation function (\ref{correlation})). Doing so, we are comparing the exact dynamics with that obtained by assuming that the map is Gaussian, and therefore fully determined by the second order moment.
\begin{figure}[ht]
\includegraphics[width=1\linewidth]{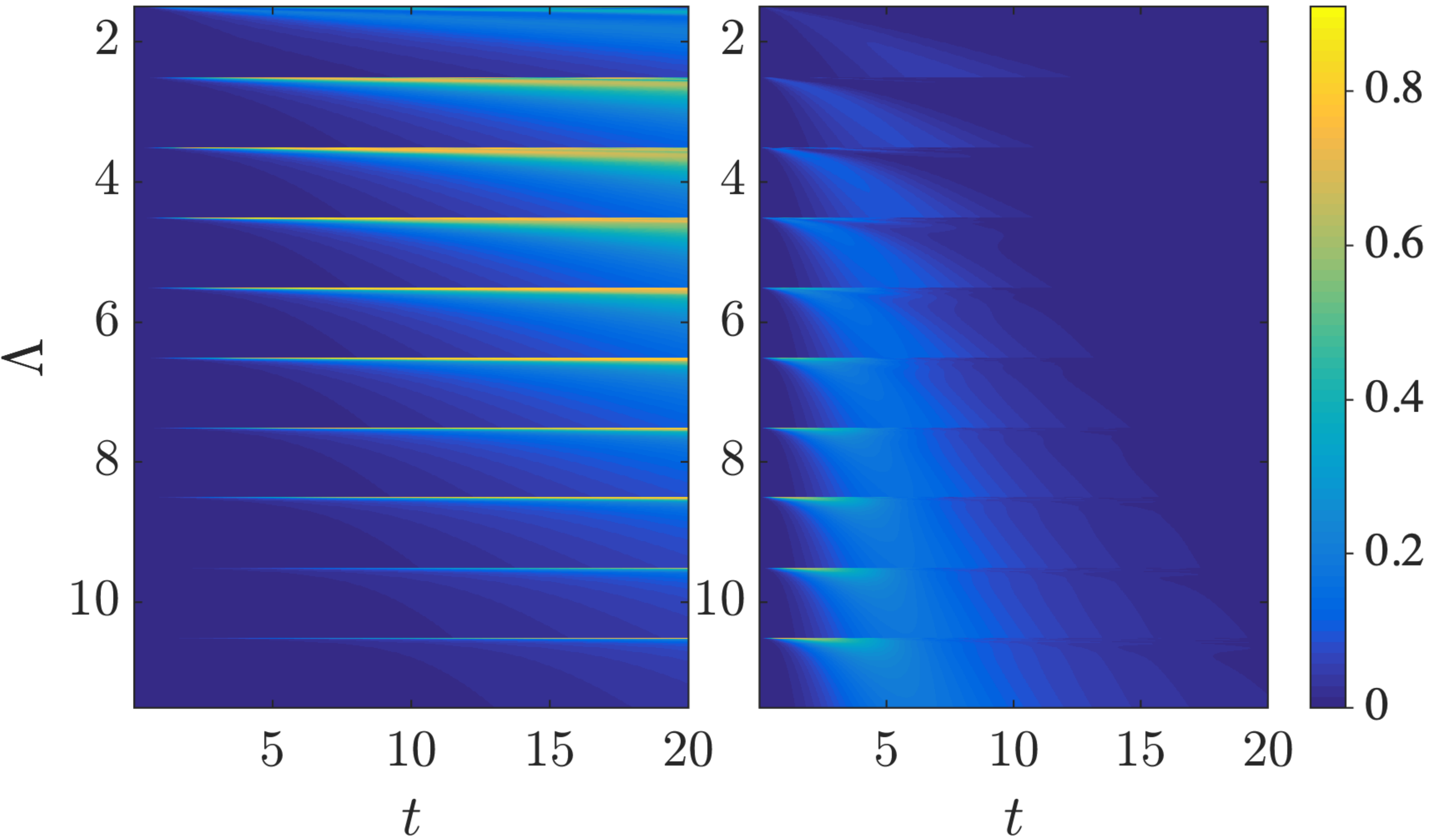}
\caption{(Color online) Density plot of the $E_{\chi}(t)=|\chi(t)-\chi^{\textmd{Gauss}}(t)|$ for $\beta=10$ (left panel) and $\beta=1$ (right panel). \label{density}}
\end{figure}
The density plot (\ref{density}) shows the error $E(t)=|\chi(t)-\chi^{\textmd{Gauss}}(t)|$ for low (left panel) and high (right panel) temperatures, for different times and anharmonicities. It can be observed in general that the Gaussian map becomes a particularly bad approximation in the regions nearby the formation of a new bound state (but not only) and at low temperatures. This behavior can be best observed in Fig. (\ref{error_gaussian}), which represents the time-averaged error
\bea
E=\frac{1}{T}\int_0^T ds D(\rho_s(t),\rho_s^{\textmd{Gauss}}(t)),
\eea
where $\rho_s(t)$ represents the exact evolution, $\rho_s^{\textmd{Gauss}}(t)$ the Gaussian-approximated evolution as given by $\chi^{\textmd{Gauss}}(t)$, and $D(A,B)=(1/2)\Ttr\{\sqrt{(A-B)^2}\}$ represents the trace distance between the two density matrices $A$ and $B$. It can be observed that the error is large for all temperatures explored except for very high values ($\beta=1$). We note that in the scale that we show it seems to increase slightly with $\Lambda$. This is a local effect which disappears, as expected, when going further to the harmonic limit. 
The lower panels show how the emergence of a new bound state (for $\Lambda=2.6$) gives rise to a strong departure of the Gaussian approximation with respect to the exact case.

\section{Conclusions}
\label{conclusions}
We have analyzed the dephasing dynamics of an impurity coupled to an environment of oscillators containing a varying degree of anharmonicity. 
We first analyze the variation of the dephasing time with the anharmonic parameter of the environment, $\Lambda$, and find the following:
\begin{itemize}
\item On a coarse-grained scale in $\Lambda$, the decoherence time increases towards both the  harmonic and anharmonic limits. In the anharmonic limit the decoherence is slow due to the few bound states in the environment. When the anharmonicity grows this effect is compensated by the fact that the environment correlation function does not decay to zero, but to a real value, the offset.  Such offset occurs due to the presence of bound states $|p\rangle$ that are highly asymmetric, i.e. for which $\langle p|x_k|p\rangle>0$ where $x_k$ is the displacement operator with respect to the potential minimum. Hence, the offset becomes negligible towards the harmonic limit, which explains the slowing down of the dephasing. 

\item On a finer scale, for a given integer number $n$ there are two different regions in $\Lambda$: type (I) regions where $\Lambda=n+1/2-\epsilon$, $\epsilon$ small, where we have an integer number of bound states $N=n$ that are all strongly bounded, i.e energetically far from the dissociation limit, and type (II) regions for $\Lambda=n+1/2+\epsilon$, $\epsilon$ small, where we have $N=n+1$ bound states and the highest energy one, $|m\rangle$, is weakly bounded. The displacement of such weakly bounded state with respect to the center of the potential is large, leading to an overlap $\langle n|x_k|n\rangle\gg 1$ that produces a particularly large offset and thus a very strong dephasing in those regions. 
\end{itemize}

Secondly, we analyze the non-Markovianity or back-flow of information, and find that it only occurs at high anharmonicities and that it is concentrated in regions (II). Moreover, the backflow is only relevant at low temperatures, which suggests that while it is indeed linked to a large anharmonicity it is nevertheless hindered by the presence of the offset. Indeed, at high temperatures such offset becomes dominant and gives rise to a dephasing dynamics that is too fast to allow for backflow. 

A similar situation occurs when considering how relevant or necessary it is to account for the non-Gaussianity of the bath. To check this we consider the correlation function of the bath and build the corresponding Gaussian map to compare it with the exact non-Gaussian one. As it turns out, for high temperatures the effect of the non-Gaussianity becomes negligible and both exact and Gaussian versions give approximately the same dynamics. Indeed, the offset is also present in the Gaussian version of the map and is so large that it dominates the decay. The non-Gaussianity, and therefore the effect of the higher order moments, starts to be important for lower temperatures, when the offset is not so relevant but still a large anharmonicity is present. In addition, the Gaussian analogue map predicts no-backflow in any regime, which reinforces the idea that the backflow is related to the anharmonicity and non-gaussian character of the bath.

Overall, this analysis unveils the presence of a variety of dynamical regimes for an impurity coupled to an anharmonic environment, and which depend not only on the number of bound states in the environment, $|m\rangle$, but also on their location within the anharmonic potential and therefore on their degree of asymmetry.  While the relevance of each of these regimes and situations remains to be determined and analyzed in each physical situation and specific model, this study is perhaps one of the first to systematically discuss how rich the dynamics of an open system may become when smoothly departing beyond the standard harmonic bath situation. In this context, concepts such as environment size effects, origin and nature of the information backflow \cite{breuer2009} and other related concepts such as divisibility \cite{rivas2009}, existence of a dynamical equation (i.e. invertibility of the map) \cite{li2018,hall2014} or validity of the weak coupling approximation may need to be revisited and further analyzed in the future. 

\begin{acknowledgements}
The authors acknowledge D. Wierichs, C. Parra, A. Recati, C. Hubig, Christiane Koch, N.O. Linden and U. Schollw\"ock for interesting discussions. This research was financially supported by DFG-grant GZ: VE 993/1-1 and supported in part by the National Science Foundation under Grant No. NSF PHY-1748958.
\end{acknowledgements}

\appendix

\section{Bound states of Morse oscillators}

We discuss in this section how to derive the bound states of the Morse potential. To this aim we will follow the derivations in \cite{morse1929} but describe in more detail some of the steps that in the original paper were only sketched. In addition we will put more emphasis on the criteria to decide whether a solution is physical or not. We start with the Hamiltonian:
	\bea
		H=\frac{p^2}{2m} + D (e^{-2\alpha x}-2e^{-\alpha x})
	\eea
	We will now do a few transformations in order to solve the stationary Schr\"odinger equation related to this Hamiltonian. First we define $x^\prime := \alpha x$. Leading to:
	\bea
	H^\prime= -\frac{\hbar^2 \alpha^2}{2m} \partial_{x^\prime}^2 +D (e^{-2x^\prime} -2e^{-x^\prime})
	\eea
	Now we rescale this to make the Hamiltonian dimensionless: $H_0=\frac{2m}{\hbar^2 \alpha^2} H^\prime$.
	Then the Hamiltonian becomes:
	\bea
	H_0=-\partial_{x^\prime}^2 + D^\prime (e^{-2x^\prime}-2e^{-x^\prime})
	\eea
	With $D^\prime=\frac{2m}{\hbar^2 \alpha^2} D:=(N+\frac{1}{2})^2$. From now on we will omit the primes.
	The bound states $\psi(x)$ for the energy $E<0$ must fulfill the following:
	\bea
	-\partial_x^2 \psi(x) + \left(N+\frac{1}{2}\right)^2 (e^{-2x}-2e^{-x}) \psi(x) -E \psi(x)=0.\nonumber
	\eea
	In order to make this a linear differential equation we substitute $z=(2N+1)\exp(-x)$. Which leads us to:
	\bea
	z^2 \partial_z^2 \psi(z) + z\partial_z \psi(z) +\left( E+\left(N+\frac{1}{2} \right)z- \frac{1}{4} z^2\right) \psi(z)=0.\nonumber
	\eea
		Now we substitute $\psi(z)=z^{b/2}\exp(-az) F(z)$. After a rather long but straight forward calculation this results in:
		\bea
		z \partial_z^2 F(z) &+& (-2az+b+1) \partial_z F(z) +\left(a^2 - \frac{1}{4}\right) z F(z)\cr
		 &+& \left(-ba -a +\left(N+\frac{1}{2}\right)\right) F(z) \cr
		&+& \left(E+\frac{b^2}{4}\right) \frac{1}{z} F(z)
		=0
		\eea
		In order to simplify this equation we choose $a=1/2$ and $b^2=-4E$. Because with this choice we obtain:
		\bea
		&&z \partial_z^2 F(z) + (b+1-z) \partial_z F(z) +\bigg(-\frac{b}{2} -\frac{1}{2} \cr
&		+&\big(N+\frac{1}{2}\big)\bigg) F(z)=0
		\eea
		This is the Laguerre equation:
		\bea
		z y^{\prime \prime} +(1+b-z) y^\prime +\lambda y=0
		\eea
	As discussed in \cite{morse1929}, one set of solutions for this equation are the Laguerre polynomials and indeed these are the only polynomial solutions. In the original paper other possible solutions are immediately discarded as unphysical. In the following we will give proof that this is the case, which justifies the approach in \cite{morse1929}. 
		Hence we start with the series ansatz:
			\bea
			y(z)=\sum_{n=0}^{\infty} a_n z^n
			\eea
			Inserting this into the Laguerre equation gives us:
			\bea
&&			\sum_{n=2}^{\infty} a_n n(n-1)z^{n-1} +\sum_{n=1}^{\infty} a_n n(1+b)z^{n-1}
		\cr
		&-& \sum_{n=1}^{\infty} a_n n z^n+ \sum_{n=0}^{\infty} \lambda a_n z^n  =0
			\eea
			Shifting the indices in the first two sums:
			\bea
			&&\sum_{n=1}^{\infty} n(a_{n+1} (n+1)-a_n)z^{n} \cr
&			+&\sum_{n=0}^{\infty} (a_{n+1} (n+1)(1+b)+\lambda a_n)z^{n}=0
			\eea
			Thus for $n=0$ we obtain the following relation between the first two coefficients:
			\bea
			a_1=\frac{-\lambda}{1+b} a_0
			\eea
			For $n>0$ we then obtain the recurrence relation:
			\bea
			a_{n+1}=\frac{-\lambda+n}{(n+1)(1+b+n)} a_n
			\eea
			Iterating this equation gives us:
			\bea
			a_n=\frac{(-\lambda)_n}{(1+b)_n n!} a_0
			\eea
			Where $(x)_n$ is the Pochhammer symbol defined as 
			\bea
			(x)_n=\Gamma(x+n)/\Gamma(x), 
			\label{poch}
			\eea
			which is equivalent to: $(x)_n=\prod_{l=0}^{n-1} (x+l)$. Please note that because of the definition of the Pochhammer symbols the series terminates if $\lambda$ is a non-negative integer. This leaves us with two cases: One is that $\lambda$ is a non-negative integer and the other one is when this is not the case. Lets first investigate the case where $\lambda$ is not an non-negative integer. We now show that in this case we do not obtain physical solutions of the Schr\"odinger equation. The argument is that if the series does not terminate, then the wavefunction will not be normalisable, and will therefore not represent a physical bound state. One can see this by splitting the series into a polynomial and a non-polynomial part and then approximate the behaviour of the non-polynomial part. The solution of the Laguerre equation is:
			\bea
			y(z)=a_0 \sum_{n=0}^{\infty} \frac{(-\lambda)_n}{(1+b)_n n!} z^n
			\eea
			First of all we note that with out loss of generality we can say that there exists an $N$ such that for all $n>N: (-\lambda +N+n)>0$. Now we split the sum as
			\bea
			y(z)=a_0 \sum_{n=0}^{N} \frac{(-\lambda)_n}{(1+b)_n n!} z^n + a_0 \sum_{n=N+1}^{\infty} \frac{(-\lambda)_n}{(1+b)_n n!} z^n\nonumber
			\eea
			We can restrict ourselves to $b>0$, since if $b<0$ then we would have a singularity in the wave function due to the substitution we made earlier. This would make it unphysical. With this the coefficients in the second sum all have the same sign. The first sum is just a polynomial and can not give us any problems, because in our substitution from earlier we have a term that scales as $\exp(-z/2)$. Thats why we abandon the polynomial part in this consideration. Now if $-\lambda>(1+b)$ then we are already done. Because then the following holds:
			\bea
&&			\abs{a_0 \sum_{n=N+1}^{\infty} \frac{(-\lambda)_n}{(1+b)_n n!} z^n} = a_0 \sum_{n=N+1}^{\infty} \abs{\frac{(-\lambda)_n}{(1+b)_n n!}} z^n\cr 
&>&  a_0 \sum_{n=N+1}^{\infty} \frac{z^n}{n!} = a_0 e^{z} - a_0 \sum_{n=0}^{N} \frac{z^n}{n!}
			\eea
			The first equality holds because we are only interested in the solution for $z$ in $[0,\infty)$. This means that in the case where $-\lambda>(1+b)$ the solution scales even stronger in $z$ then $\exp(z)$ and thus it is unphysical, because neither our factor of $\exp(-z/2)$ nor any polynomial can compete with that. So now lets see how this goes for $-\lambda<(1+b)$. In this case this is a bit more difficult. First we need to note that $\lim_{n\rightarrow\infty} (-\lambda+n)/(1+b+n)=1$. This means for all $\e>0$ there exists an $N^\prime$ such that for all $n\geq N^\prime$:
			\bea 
			\abs{\frac{-\lambda+n}{1+b+n} -1}<\e 
			\eea
			From which follows that for $n\geq N^\prime: \frac{-\lambda+n}{1+b+n} > 1-\e$. With this we can investigate the non-polynomial part further:
			\bea
&&			a_0\sum_{n=N+1}^{\infty} \frac{(-\lambda)_n}{(1+b)_n n!} z^n= a_0\sum_{n=N+1}^{N^\prime} \frac{(-\lambda)_n}{(1+b)_n n!} z^n\nonumber \cr &+&  a_0\sum_{n=N^\prime+1}^{\infty} \frac{(-\lambda)_n}{(1+b)_n n!} z^n
			\eea
			Now lets just look at the non-polynomial part again:
			\bea
&&			\sum_{n=N^\prime+1}^{\infty} \frac{(-\lambda)_n}{(1+b)_n n!} z^n \cr
&=& c_{N^\prime} \sum_{n=N^\prime+1}^{\infty} \left(\prod_{l=0}^{n-1} \frac{-\lambda+N^\prime +l}{1+b+N^\prime+l}\right) \frac{z^n}{n!}\cr
			&>& c_{N^\prime}\sum_{n=N^\prime+1}^{\infty} \left(\prod_{l=0}^{n-1} (1-\e)\right) \frac{z^n}{n!}\cr
			&=& c_{N^\prime}\, e^{(1-\e)z} - c_{N^\prime}\sum_{n=0}^{N^\prime} (1-\e)^n \frac{z^n}{n!}
			\eea
			Here $c_{N^\prime} :=\frac{(-\lambda)_{N^\prime}}{(1+b)_{N^\prime}}$ and $\e$ can be chosen to be any positive real number. Thus if one chooses for example $\e=1/4$ one can see that the wave function is not normalisable. Again the second term does not make it normalisable again, since it is just a polynomial. With this extensive proof we showed that the only physical solutions are obtained for $\lambda$ being a non-negative integer.
			With this we obtain an equation for $b$ and for the energy since $b^2/4=-E$:
			\bea
			n&=&N-\frac{b}{2} \\
			b&=&2N-2n\\
			E_n&=&-(N - n)^2
			\eea
			Here $\lambda \rightarrow n$ is a non-negative integer. Remember that $b$ is only allowed to be positive, because otherwise either the wave functions or the probability density will have a singularity at $z=0$. This means that if $N$ is an natural number it is the number of bound states, if not then the integer part of $N+1$ is the number of bound states. The wave function is then given by:
			\bea
			\psi_n(z)&=& z^{2N-2n} e^{-z/2} y_n^{2N-2n}(z)\\
			&=& z^{2N-2n} e^{-z/2} \sum_{m=0}^{n} \frac{(-n)_m}{(2N-2n+1)_m m!} z^m\\ \nonumber
			&=& z^{2N-2n} e^{-z/2} \sum_{m=0}^{n} (-1)^m \binom{n}{m} \frac{z^m \Gamma(2N-2n+1)}{\Gamma(2N-2n+1+m)} \nonumber
			\eea
			With proper normalisation this becomes:
			\bea
			\psi_n(z)&=&N_n z^{2N-2n} e^{-z/2} \sum_{m=0}^{n} (-1)^m
\binom{n}{m}  \cr			&\times&\frac{z^m }{\Gamma(2N-2n+1+m)}.
\label{psi}
			\eea
Here we have defined $z$ and $N_n$ as follows:
			\bea
			z&=&(2N+1)\exp(-x) \\
			N_n&=&\sqrt{\frac{(2N-2n) \Gamma(v^N_{n})}{n!}}
			\eea
			Note that all these equations are still for a dimensionless Hamiltonian. When we reintroduce the parameters we rescaled the Hamiltonian with in the beginning we get:
			\bea
			E_n&=&-\frac{\hbar^2 \alpha^2}{2m}(N - n)^2\\
			&=&-\frac{\hbar^2 \alpha^2}{2m}\!\left(\Lambda^2 +\left(n+\frac{1}{2}\right)^2 - 2\Lambda \left(n+\frac{1}{2}\right)\right).\nonumber
			\eea
			Where $\Lambda$ is defined as $\Lambda := N+\frac{1}{2}$. Now for a given Morse Oscillator we want the part that is proportional to $n+\frac{1}{2}$ to be the harmonic part thus it should be $\hbar \w (n+\frac{1}{2})$. So now we change the parameters of the Morse Oscillator from $\alpha$ and $D$ to $\w$ and $\Lambda$:
			\bea
			\alpha &:=& \sqrt{\frac{m \w}{\hbar \Lambda}} \\
			D &:=& \frac{\hbar \w \Lambda}{2}
			\eea 
			As mentioned earlier we obtain a set $\{\w_k, g_k\}$ by discretizing the spectral density. The $\w$ in the above equations corresponds to the $\w_k$ that we obtain by the discretization. For the implementation of the Morse environment we choose $\Lambda$ to be constant for all oscillators, thus they all have the same number of bound states. 

\section{Matrix elements of the position operator}
\label{matrix_elem}
In our description, it is highly convenient to write the Morse Hamiltonian in its diagonal form. This in turn means that the interaction Hamiltonian should also be expressed in such basis. To this aim, it is necessary to write the matrix element of the position operator in such basis, a computation that as we will see in the following is not at all trivial. To proceed with it, we shall make use of the wave function of the bound states, as computed in the previous appendix. In the following, we consider the same strategy as in Ref. \cite{lima2005}. That is we compute them as:
	\bea
	\bra{n}x\ket{m}=\lim_{\eta\rightarrow0} \text{Im}\left(\frac{d}{d\eta}\bra{n}\exp(i \eta x)\ket{m}\right)
	\eea
	Here $x$ is still supposed to be dimensionless, thus the rescaling of the last section is still in place. Actually, computing $\bra{n}\exp(i \eta x)\ket{m}$ is pretty straight forward apart from the fact that one has to use the gaussian hypergeometric theorem and the Saalsch\"utz theorem. Those will both be stated when we use them. Nevertheless, the most delicate part of the computation is taking the limit after the derivative. Since this important step was not detailed in \cite{lima2005}, we will present it here in more detail.\\
	Let us first consider that $\exp(i\eta x)=(2N+1)^{i\eta}z^{-i\eta}$, such that
	\bea
	\bra{n}\exp(i\eta x)\ket{m}&=&\int_{-\infty}^{\infty}dx \psi_n^*(z) \psi_m(z) (2N+1)^{i\eta} z^{-i\eta}\cr
	&=& \int_0^{\infty} dz \psi_n^*(z) \psi_m(z) (2N+1)^{i\eta} z^{-i\eta-1}\nonumber
\eea
	Then, considering the wave functions (\ref{psi}) we can now compute:
\bea
&&		\bra{n}\exp(i\eta x)\ket{m}=(2N+1)^{i\eta} N_n N_m\! \sum_{\np, \mpr=0}^{n,m} \binom{n}{\np} \binom{m}{\mpr} \nonumber\\
	&\times& (-1)^{\np+\mpr}\frac{1}{\Gamma(u^N_{nn}+\np +1)\Gamma(u^N_{mm}+\mpr+1)}\cr
	&\times& \int_{0}^{\infty} dz e^{-z} z^{u^N_{nm}+\np+\mpr-i\eta-1},
	\eea
where we have defined 
			\bea
			u^N_{nm}&=&2N-n-m,\\\nonumber
			v^N_{n}&=&2N-n+1.
			\eea
	Using the definition of the $\Gamma$-function: $\Gamma(z)=\int_{0}^{\infty} dt e^{-t} t^{z-1}$ one obtains:
	\bea
&&	\bra{n}\exp(i\eta x)\ket{m}=(2N+1)^{i\eta} N_n N_m \sum_{n'=0}^{n}\frac{\Sigma(m,n,\eta)}{\Gamma(u^N_{nn}+\np +1)},\cr
&&
	\eea
	where we have defined a function that contains all terms that depend on $\mpr$
	\bea
&&	\Sigma(m,n,\eta):=
\sum_{\mpr=0}^{m}\binom{m}{\mpr}(-1)^\mpr \frac{\Gamma(u^N_{nm}+\np+\mpr-i\eta)}{\Gamma(u^N_{mm}+\mpr+1)}.\cr
&&
\label{gamma}
\eea
We now we consider that 
	\bea
	(-m)_\mpr=\prod_{l=0}^{\mpr-1} (-m+l)=
	(-1)^\mpr \frac{m!}{(m-\mpr)!}
	\eea
in order to re-express Eq. (\ref{gamma}) in terms of Pochhammer symbols (\ref{poch}),
	\bea
&&	\Sigma(m,n,\eta):=
	\sum_{\mpr=0}^{m} \frac{\Gamma(u^N_{nm}+\np-i\eta)}{\Gamma(u^N_{mm}+1)}\nonumber\\
	&\times& \frac{(u^N_{nm}+\np-i\eta)_\mpr (-m)_\mpr}{(u^N_{mm}+1)_\mpr \mpr!}. 
	\eea
This allows us to perform the sum over $\mpr$ by using the gaussian hypergeometric theorem, which states (see \cite{lima2005} section 3.1)
	\bea
	\sum_{\mpr=0}^{\infty} \frac{(a)_\mpr (b)_\mpr}{(c)_\mpr \mpr!}=\frac{\Gamma(c)\Gamma(c-a-b)}{\Gamma(c-a)\Gamma(c-b)}	
	\eea
		Note that $(-m)_\mpr$ becomes 0, when $\mpr>m$, which allows us to apply the  theorem even though it is meant for summation up to infinity. When using this we obtain:
		\bea
		\Sigma(n,m,\eta)=\frac{\Gamma(u^N_{nm}+\np-i\eta)\Gamma(n-\np+1+i\eta)}{\Gamma(n-m+1-\np +i\eta) \Gamma(v^N_{m})}\nonumber
		\eea
		Thus the matrix element becomes:
		\bea
&&		\bra{n}\exp(i\eta x)\ket{m}=\frac{(2N+1)^{i\eta}N_n N_m}{\Gamma(v^N_{m})} \sum_{\np=0}^{n} \binom{n}{\np} (-1)^{\np}\cr &\times&\frac{\Gamma(u^N_{nm}+\np-i\eta)\Gamma(n-\np+1+i\eta)}{\Gamma(n-m+1-\np +i\eta) \Gamma(u^N_{nn}+\np+1)}
		\eea
			Now we want to organise this expression into two terms, where one contains everything depending on $\np$. Also we want to have the things that depend on $\np$ to be formulated in terms of Pochhammer symbols again, so that we can ultimatly use Saalsch\"utz theorem. First we have to note the following:
			\bea
			\Gamma(a)= 
			(-1)^{\np} (-a+1)_{\np} \Gamma(a-\np)  
			\eea
			With this we obtain:
			\bea
			\bra{n}\exp(i\eta x)\ket{m}&=&A_{mn}(\eta) \Gamma_{mn}(\eta)
			\eea
			Where $A_{mn}(\eta)$ and $\Gamma_{mn}(\eta)$ are defined as follows:
			\bea
&&			A_{mn}(\eta)=\frac{(2N+1)^{i\eta}N_n N_m}{\Gamma(v^N_{m})} \frac{\Gamma(u^N_{nm}-i\eta)\Gamma(n+1+i\eta)}{\Gamma(n-m+1+i\eta)\Gamma(u^N_{nn}+1)}\cr
&&			\Gamma_{mn}(\eta)=
_3F_2(-n+m-i\eta,u^N_{nm}-i\eta,-n;u^N_{nn}+1,\nonumber\\
			&&-n-i\eta;1)
			\label{AG}
			\eea
			Where $_3F_2$ is a generalised hypergeometrical function as can also be seen in \cite{lima2005} section 3.1 eq. 18 and \cite{bateman1953} chapter 4.
			The derivative can now be computed as:
			\bea
&&			\frac{d}{d\eta}\bra{n}\exp(i\eta x)\ket{m}=\left(\frac{d}{d\eta} A_{mn}\right)\!(\eta) \,\Gamma_{mn}(\eta)\cr
			&+&A_{mn}(\eta)\left(\frac{d}{d\eta}\Gamma_{mn}\right)\!(\eta)
			\eea
			Up to now we went pretty much along the lines of \cite{lima2005} apart from expressing things in a much more detailled manner of course. Taking this derivative and the limit afterwards is the most problematic part of this computation and since this step is skipped in \cite{lima2005} we are going to do this in all detail here.\\
			Let us distinguish two cases. The offdiagonal elements namely $m>n$ and the diagonal elements. Note that the condition $m>n$ does not imply a loss of generality since x is hermitian. First we take a look at the offdiagonal elements. For $m>n$ one can see that in the denominator of $A_{mn}(\eta)$ there is a term $\Gamma(n-m+1+i\eta)$ which approaches a singularity in the limit $\eta\rightarrow0$. Thus  $A_{mn}(\eta)$ becomes 0 in the limit. If $\frac{d}{d\eta}\Gamma_{nm}(\eta)$ does not have a diverging term in the limit this means, that we only have to take into account the term containing the derivative of $A_{mn}(\eta)$. By looking at the form of $\Gamma_{mn}(\eta)$ we can already see that taking a derivative and sending $\eta$ to 0 will not give us something divergent. Thus we only care for the derivative of $A_{mn}(\eta)$.
			\bea
			\lim_{\eta\rightarrow0}\frac{d}{d\eta}A_{mn}(\eta)&=&\frac{(2N+1)^{i\eta}N_n N_m \Gamma(n+1) \Gamma(u^N_{nm})}{\Gamma(v^N_{m})\Gamma(u^N_{nn}+1)} \cr
			&\times&(-i) \lim_{\eta\rightarrow0} \frac{\psi(n-m+1+i\eta)}{\Gamma(n-m+1+i\eta)}
			\eea
			where $\psi(x):=\frac{d}{dx} \ln(\Gamma(x))=\frac{\frac{d}{dx}\Gamma(x)}{\Gamma(x)}$ is the digamma function. 
			Here we skipped a few steps, but they are just taking the derivative and realising, that in the limit $\eta\rightarrow0$ only this term survives, because in all the other terms there is no singularity in the enumerator. In order to obtain this limit one can do a nice trick. Namely one can use the following two identities(\cite{bateman1953} chapter 1.2 eq. 6 and chapter 1.7 eq. 11):
			\bea
			\Gamma(z)\Gamma(1-z)&=&\pi \csc(\pi z)\\
			\psi(1-z)-\psi(z)&=&\pi \cot(\pi z).
			\eea
			With the use of these equations, the limit becomes:
			\bea
			\lim_{\eta\rightarrow0} \frac{\psi(n-m+1+i\eta)}{\Gamma(n-m+1+i\eta)}
			&=&-\cos(\pi(n-m+1))\Gamma(m-n) \nonumber \\
			&=& (-1)^{m-n} \Gamma(m-n)
			\eea
				With this we obtain(for $m>n$):
				\bea
				\lim_{\eta\rightarrow0}\frac{d}{d\eta}A_{mn}(\eta)&=& i(-1)^{m-n+1} \frac{N_n N_m \Gamma(n+1) }{\Gamma(v^N_{m})\Gamma(u^N_{nn}+1)}\nonumber \\
				&\times&  \Gamma(u^N_{nm})\Gamma(m-n)
				\eea
				Thus the only thing that remains to be done in the offdiagonal case is to take the limit for $\Gamma_{mn}(\eta)$. If one takes this limit one obtains:
				\bea
				\lim_{\eta\rightarrow0} \Gamma_{mn}(\eta)=_3F_2(-n+m,u^N_{nm},-n;u^N_{nn}+1,-n;1)\nonumber
				\eea
				A generalized hypergeometric function $_3F_2$ is Saalsch\"utzian if its parameters are in the following form and the Saalsch\"utz theorem states that (see \cite{bateman1953} chapter 4.4):
				\bea
				_3F_2(a,b,-n;c,1+a+b-c-n,1)&=&\frac{(c-a)_n (c-b)_n}{(c)_n (c-a-b)_n}.
			\cr	
			&&
			\label{Saa_th}
				\eea
				In the limit $\eta \rightarrow 0$ the parameters for our generalised hypergeometric function are of exactly that form, thus:
				\bea
				\lim_{\eta\rightarrow0} \Gamma_{mn}(\eta)&=& \frac{(-m)_n (-2N+m)_n}{(-n)_n (-2N-n)_n}
				\eea
Here, we have considered that 
				\bea
				(-x)_n=\prod_{l=0}^{n-1}(-x+j)
				=(-1)^n \frac{\Gamma(x+1)}{\Gamma(x-n+1)}.
				\label{definition}
				\eea
		With this we obtain:
		\bea
		\lim_{\eta\rightarrow0} \Gamma_{mn}(\eta)&=& \frac{m! \Gamma(v^N_{m}) \Gamma(u^N_{nn}+1)}{n!\Gamma(v^N_{n})\Gamma(u^N_{nm}+1) \Gamma(m-n+1)}\nonumber
		\eea
		Now we can finally obtain the matrix elements for the offdiagonal term with $m>n$:
		\bea
&&		\bra{n}x\ket{m}=\frac{(-1)^{m-n+1}m!N_n N_m \Gamma(u^N_{nm})\Gamma(m-n)}{\Gamma(v^N_{n})\Gamma(u^N_{nm}+1) \Gamma(m-n+1)}\cr
		&=& \frac{2(-1)^{m-n+1}}{u^N_{nm}(m-n)}
		 \sqrt{\frac{m!(N-n)(N-m)\Gamma(v^N_{m})}{n! \Gamma(v^N_{n})}} \nonumber
		\eea
		The last two lines show our result for the offdiagonal elements. Now for the diagonal elements $A_{mn}(\eta)$ does not have a singularity in the denominator, thus both terms contribute. Let us first compute the contribution from $A_{nn}(\eta)$:
		\bea
		\lim_{\eta\rightarrow0} \frac{d}{d\eta} A_{nn}(\eta)&=& i\ln(2N+1) - i\psi(u^N_{nn}) \cr
		&+&i\psi(n+1) -i \psi(1)
		\eea
		Here we have not detailed the derivation, which basically consists on taking the derivative of each term of $A_{nn}(\eta)$ in Eq. (\ref{AG}) and then taking the limit. Please note that the derivation of $\lim_{\eta\rightarrow0} \Gamma_{nm}(\eta)$ was not relying on $m>n$ so it also holds here, and thus $\lim_{\eta\rightarrow0} \Gamma_{nn}(\eta)=1$. So now comes the tricky part. In order to obtain a nice form for the derivative of $\Gamma_{nn}(\eta)$ we would like to use the Saalsch\"utz theorem of Eq. (\ref{Saa_th}). But if we are not in the limit where $\eta \rightarrow 0$ our $_3F_2$ is not Saalsch\"utzian. This is where we use a trick. We define another $\tilde{\Gamma}_{nn}(\eta)$, that is Saalsch\"utzian and that has the same derivative in the limit $\eta\rightarrow0$. Define $\tilde{\Gamma}_{nn}(\eta)$ as follows:
		\bea
		\tilde{\Gamma}_{nn}(\eta):=_3F_2(-i\eta,u^N_{nn}-i\eta,-n;u^N_{nn}+1-i\eta,-n-i\eta;1).\nonumber
		\eea
		First we have to check that in the limit of $\eta \rightarrow0$ this indeed has the same derivative as $\Gamma_{nn}(\eta)$. So lets take a look at how this is expressed in terms of Pochhammer symbols:
		\bea
		\tilde{\Gamma}_{nn}(\eta)=\sum_{\np=0}^{n} \frac{(-i\eta)_{\np} (u^N_{nn}-i\eta)_{\np} (-n)_{\np}}{(u^N_{nn}+1-i\eta)_{\np} (-n-i\eta)_{\np}}.
		\eea
		Because of the term $(-i\eta)_{\np}$ this is proportional to $\eta$, so the only term that contributes to the derivative in the limit of $\eta\rightarrow0$ is the one where one takes the derivative of $(-i\eta)_{\np}$ (in other words, the derivative of the other terms with $\eta$ will always be multiplied by $(-i\eta)_{\np}$ and therefore vanish in the limit $\eta\rightarrow 0$). This leads to:
		\bea
		\lim_{\eta\rightarrow0}\frac{d}{d\eta}\tilde{\Gamma}_{nn}(\eta)&=&\lim_{\eta\rightarrow0}\sum_{\np=0}^{n} \frac{\left(\frac{d}{d\eta}(-i\eta)_{\np}\right) (u^N_{nn}-i\eta)_{\np} (-n)_{\np}}{(u^N_{nn}+1-i\eta)_{\np} (-n-i\eta)_{\np}}\cr
		&=&\sum_{\np=0}^{n} \frac{\lim_{\eta\rightarrow0}\left(\frac{d}{d\eta}(-i\eta)_{\np}\right) (u^N_{nn})_{\np} (-n)_{\np}}{(u^N_{nn}+1)_{\np} (-n)_{\np}}\nonumber
		\eea
		Now let us compare this to $\Gamma_{nn}(\eta)$:
		\bea
		\Gamma_{nn}(\eta)=\sum_{\np=0}^{n} \frac{(-i\eta)_{\np} (u^N_{nn}-i\eta)_{\np} (-n)_{\np}}{(u^N_{nn}+1)_{\np} (-n-i\eta)_{\np}}
		\eea
		With the same argument as for $\tilde{\Gamma}_{nn}(\eta)$ we obtain:
		\bea
&&		\lim_{\eta\rightarrow0}\frac{d}{d\eta}\Gamma_{nn}(\eta)\cr
		&=&\sum_{\np=0}^{n} \frac{\lim_{\eta\rightarrow0}\left(\frac{d}{d\eta}(-i\eta)_{\np}\right) (u^N_{nn})_{\np} (-n)_{\np}}{(u^N_{nn}+1)_{\np} (-n)_{\np}}\cr
		&=& \lim_{\eta\rightarrow0}\frac{d}{d\eta}\tilde{\Gamma}_{nn}(\eta).
		\eea
		Now we use the Saalsch\"utz theorem to compute $\tilde{\Gamma}_{nn}(\eta)$:
		\bea
&&		\tilde{\Gamma}_{nn}(\eta)=\frac{(u^N_{nn}+1)_n (1)_n}{(u^N_{nn}+1-i\eta)_n (1+i\eta)_n} \\
		&=&\frac{\Gamma(v^N_{n})\Gamma(1+n)\Gamma(u^N_{nn}+1-i\eta)\Gamma(1+i\eta)}{\Gamma(u^N_{nn}+1)\Gamma(v^N_{n}-i\eta)\Gamma(1+i\eta+n)}\nonumber
		\eea
		Thus:
		\bea
		\lim_{\eta\rightarrow0}\frac{d}{d\eta}\tilde{\Gamma}_{nn}(\eta)&=&-i\psi(u^N_{nn}+1)+i\psi(1)\nonumber\\&+&i\psi(v^N_{n})-i\psi(1+n)
		\eea
		Also note that $\lim_{\eta\rightarrow0}A_{nn}(\eta)=1$. With this we can compute the diagonal matrix elements:
		\bea
		\bra{n}x\ket{n}&=&\text{Im}\left(\lim_{\eta\rightarrow0}\frac{d}{d\eta}A_{nn}(\eta) + \lim_{\eta\rightarrow0} \frac{d}{d\eta} \Gamma_{nn}(\eta)\right)\cr
		&=& \ln(2N+1) -\psi(u^N_{nn}) -\psi(u^N_{nn}+1)\nonumber\\
		&+&\psi(v^N_{n}) \label{eq:3.1}
		\eea
		For the sake of completeness let us state the offdiagonal elements for $m>n$ here again:
		\bea
		\bra{n}x\ket{m}&=&\frac{2(-1)^{m-n+1}}{u^N_{nm}(m-n)}\\
		&\times& \sqrt{\frac{m!(N-n)(N-m)\Gamma(v^N_{m})}{n! \Gamma(v^N_{n})}} \nonumber \label{eq:3.2}
		\eea

		Here we still refer to the dimensionless $x$ thus in order to obtain the matrix elements of $b^\dagger+b$ one only has to rescale the ones from (\ref{eq:3.1}) and (\ref{eq:3.2}) by a factor $\sqrt{2/\Lambda}$, namely
		\bea
		b^\dagger + b =\sqrt{\frac{2m\w}{\alpha \hbar}} x = \sqrt{\frac{2}{\Lambda}} x. \label{eq:3.86}
		\eea
		
\label{decay_spin}

\bibliography{/Users/ines.vega/Dropbox/Bibtexelesdrop2}

\end{document}